\documentclass[conference]{IEEEtran}
\IEEEoverridecommandlockouts
\usepackage{cite}
\usepackage{amsmath,amssymb,amsfonts}
\usepackage{algorithmic}
\usepackage{graphicx}
\usepackage{textcomp}
\usepackage{xcolor}
\usepackage{hyperref}
\usepackage{bm}

\def\BibTeX{{\rm B\kern-.05em{\sc i\kern-.025em b}\kern-.08em
    T\kern-.1667em\lower.7ex\hbox{E}\kern-.125emX}}
\begin{document}

\title{Uplink OFDM Channel Prediction with Hybrid CNN-LSTM for 6G Non-Terrestrial Networks \\
}

\author{\IEEEauthorblockN{Bruno De Filippo\IEEEauthorrefmark{1}, Carla Amatetti\IEEEauthorrefmark{1}, Alessandro Vanelli-Coralli\IEEEauthorrefmark{1}}
\IEEEauthorblockA{\IEEEauthorrefmark{1}Department of Electrical, Electronic, and Information Engineering (DEI), Univ. of Bologna, Bologna, Italy}
\{bruno.defilippo, carla.amatetti2, alessandro.vanelli\}@unibo.it}

\maketitle

\begin{abstract}
Wireless communications are typically subject to complex channel dynamics, requiring the transmission of pilot sequences to estimate and equalize such effects and correctly receive information bits. This is especially true in 6G non-terrestrial networks (NTNs) in low Earth orbit, where one end of the communication link orbits around the Earth at several kilometers per second, and a multi-carrier waveform, such as orthogonal frequency division multiplexing (OFDM), is employed. To minimize the pilot overhead, we remove pilot symbols every other OFDM slot and propose a channel predictor to obtain the channel frequency response (CFR) matrix in absence of pilots. The algorithm employs an encoder-decoder convolutional neural network and a long short-term memory layer, along with skip connections, to predict the CFR matrix on the upcoming slot based on the current one. We demonstrate the effectiveness of the proposed predictor through numerical simulations in tapped delay line channel models, highlighting the effective throughput improvement. We further assess the generalization capabilities of the model, showing minimal throughput degradation when testing under different Doppler spreads and in both line of sight (LoS) and non-LoS propagation conditions. Finally, we discuss computational-complexity-related aspects of the lightweight hybrid CNN-LSTM architecture.
\end{abstract}


\section{Introduction}\label{ch:1_intro}
In the last decade, artificial intelligence (AI) algorithms, such as neural networks (NNs), have been shown providing significant benefits to the telecommunications field, either in terms of computational complexity \cite{bib:neuralRX} and/or communications-related metrics \cite{bib:SPAWC}. This has been recognized by the 3rd generation partnership project (3GPP), leading to several technical reports and specifications of AI-related aspects, including the application of AI to the radio access network \cite{bib:tr38.843}. Among the proposed use cases, channel prediction has attracted particular interest as a means to obtain information on the status of the propagation channel when its estimation is not practical. This is the case of non-terrestrial networks (NTNs), one of the pillars of the 6th generation (6G) of wireless communication standards \cite{bib:6GNTN}, typically characterized by dynamic channel conditions including time-varying delays and Doppler frequency shifts. In this context, channel prediction has been proposed to counter the channel aging effect, \textit{i.e.}, the drop in accuracy of estimated channel information due to the change of the propagation conditions (\textit{e.g.}, a low Eart orbit (LEO) satellite's movement), which has been proven to hinder the performance of satellite communication systems \cite{bib:aging}. A long short-term memory (LSTM)-based NN was employed in \cite{bib:channelAgingSat} to bring up to date channel estimates relayed by user equipments (UEs) in time division duplexing NTNs. This has a particular relevance for NTNs, where channel estimates feedbacks are typically necessary to compute beamforming vectors in multiple input multiple output (MIMO) systems. Channel aging in NTNs was also tackled in \cite{bib:SPAWC} and \cite{bib:upToDown} using a lightweight deep NN and a convolutional NN (CNN) with LSTM, respectively. On the opposite, the authors of \cite{bib:SCP} trained an LSTM to predict a time series of MIMO channel matrices based on the most recent estimates, reducing the normalized mean squared error (NMSE) of the predictions by 7 dB with respect to a recurrent NN. A similar approach was employed in \cite{bib:precoding}, with the predictor consisting of a CNN with LSTM and the channel data being augmented with a variational autoencoder. Finally, the authors in \cite{bib:attentionPred} tackled the prediction of single carrier channel coefficients by proposing a CNN-LSTM with an attention layer, resulting in an NMSE improvement with respect to the model introduced in \cite{bib:SCP}.

As shown by the literature, channel prediction has great potential in NTNs. However, most of the works in the literature focus on the NMSE as a comparison metric, neglecting to assess the impact of the predictors on key communication metrics. In our previous work \cite{bib:icmlcn}, we provided a preliminary evaluation of physical layer metrics, \textit{i.e.}, bit error rate (BER), block error rate (BLER), and throughput (TP), in 6G NTN systems with downlink channel prediction, assuming orthogonal frequency division multiplexing (OFDM) as a waveform. In particular, a convolution-based encoding-decoding NN was used to predict the upcoming OFDM channel frequency response (CFR) matrix from the current estimates. Building up from \cite{bib:icmlcn}, in this paper we:
\begin{itemize}
    \item Propose a novel lightweight channel predictor based on a hybrid CNN-LSTM with skip connections, leveraging the pattern extraction capabilities of convolution-based encoder-decoder architectures and predicting time series of low-level features with an LSTM layer;
    \item Assess the throughput gains achieved by the proposed algorithm in a LEO-based NTN considering realistic channel models, \textit{e.g.}, NTN-TDL-C, etc \cite{bib:tr38.811};
    \item Evaluate the generalization capabilities of the pre-trained model when tested on different user equipment (UE) speeds and channel models.
\end{itemize}

\section{System model}\label{ch:2_systemModel}
We consider a set of UEs on ground, equipped with a 6G transceiver and characterized by various degrees of mobility, communicating in uplink with a gNodeB (gNB) hosted on a LEO satellite. The UEs are assumed to be able to pre-compensate the Doppler shift and keep track of the Doppler rate resulting from the mobility of the considered NTN node, with just a residual frequency synchronization error affecting the received cyclic prefix OFDM (CP-OFDM) waveform. The transmission chain at the UEs employs a low density parity check (LDPC) channel encoder, a bit interleaver, an $M$-ary quadrature amplitude modulation (M-QAM) mapper, and a CP-OFDM multiplexer. Focusing on a single UE, we assume that $N_{SC}$ subcarriers have been allocated for uplink transmission. The resource grid $\mathbf{X} \in \mathbb{C}^{N_{SC}\times N_{sym}}$ is thus generated. We assume that, within a slot, $M_\pi$-QAM pilot symbols are inserted over each subcarrier of the resource grid at OFDM symbol indices $\mathcal{I}_\pi^{(slot)}$. However, adopting the same strategy as in \cite{bib:icmlcn}, we reduce the pilot overhead by removing all pilot symbols from every other slot, \textit{i.e.}, half of the slots only contains data symbols; thus, we here limit the analysis to $N_{sym}=2N_{sym}^{(slot)}$ OFDM symbols, with $N_{sym}^{(slot)}=14$ representing the number of OFDM symbols per slot. We note that this approach results in a theoretical peak throughput uplift by a factor $N_{sym}/(N_{sym}-\left|\mathcal{I}_\pi^{(slot)}\right|)$, where $\left|\mathcal{I}_\pi^{(slot)}\right|$ represents the number of OFDM pilot symbols per slot. The CP-OFDM waveform carrying the resource grid is transmitted by the UE and propagates through a frequency-selective channel, modeled as a tapped delay line (TDL) with the following $N_{taps}$-paths channel impulse response:
\begin{equation}\label{eqn:impulseResponse}
    h(t, \tau) = e^{j2\pi\epsilon_Dt}\sum_{n=1}^{N_{taps}}h_n(t)\delta(\tau-\tau_n),
\end{equation}
\begin{figure}[t]
    \centerline{\includegraphics[width=8.4cm]{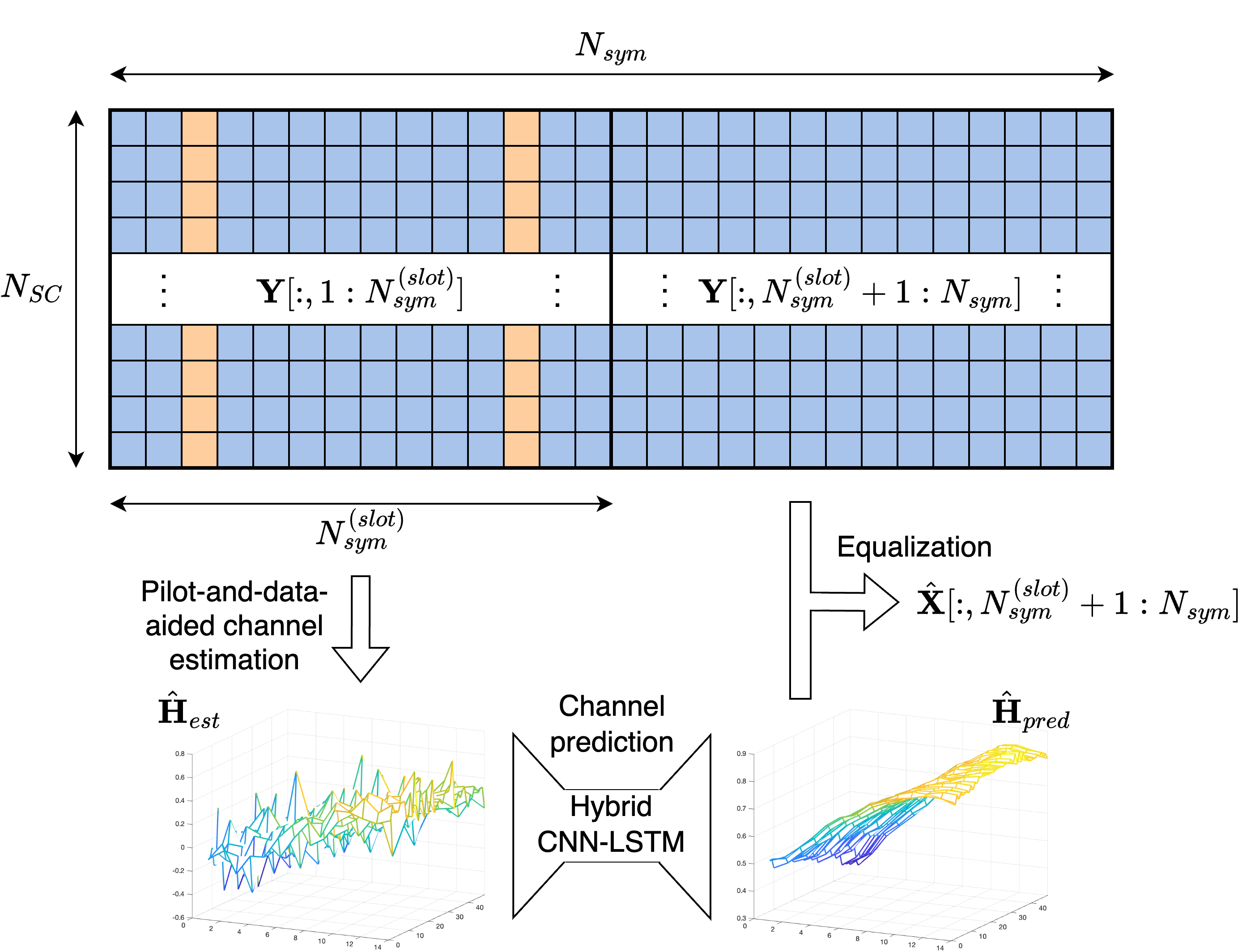}}
    \caption{Channel prediction framework (blue and orange squares correspond to data and pilot symbols, respectively).}
    \label{fig:framework}
\end{figure}
where $\epsilon_D\sim\mathcal{N}(0, \sigma_D)$ is the residual frequency synchronization error component, modeled as a zero-mean Gaussian random variable with standard deviation $\sigma_D$, $h_n(t)$ and $\tau_n$ represent the value and delay (typically assuming $\tau_1 = 0$) of the $n$-th tap, respectively, and $\delta(\cdot)$ represents the Dirac's delta function. At the receiver, additive white Gaussian noise (AWGN) corrupts the waveform; then, the cyclic prefix is removed from the signal and the received symbols are de-multiplexed. Representing with $\mathbf{H} = FFT[h(t, \tau)]$ the CFR matrix, with $FFT$ being the fast Fourier transform algorithm, the OFDM resource grid at the receiver can be expressed as:
\begin{equation}
    \mathbf{Y} = \mathbf{H}\odot\mathbf{X} + \mathbf{W},
\end{equation}
where $\mathbf{W}$ represents the effect of AWGN on the OFDM grid (with $N_0$ representing the noise power spectral density), and $\odot$ is the Hadamard operator. To obtain an estimate $\hat{\mathbf{H}} \in \mathbb{C}^{N_{SC}\times N_{sym}}$ for the CFR matrix, least square (LS) channel estimation is first performed on the received pilot symbols as $\hat{\mathbf{H}}_\pi=\mathbf{X}[:, \mathcal{I}_\pi^{(slot)}]^{-1}\mathbf{Y}[:, \mathcal{I}_\pi^{(slot)}]$ \cite{bib:channelEst}; then, such estimates are interpolated over the slot duration to cover the temporal span of the resource grid. Clearly, with the chosen system model, interpolation should cover not only the data symbols positions over the first slot, but also all of the symbols locations over the second slot, where no pilots are present; thus, we rely on channel prediction instead to improve the channel estimates accuracy (Figure \ref{fig:framework}). First, the data symbols on the first slot are equalized with the obtained channel estimates and demapped; then, the resulting data bits are remapped over M-QAM symbols, which are used to perform LS over the entire first slot, resulting in an overall data-and-pilot-aided estimation scheme. On the one hand, when symbols are correctly demapped, the second LS step filters out channel estimation inaccuracies due to interpolation; on the other hand, however, demapping errors lead to quantized inaccuracies on the estimated channel matrix, \textit{e.g.}, an error on a 4-QAM data symbol may result in a $\pi/2$ phase shift of the corresponding channel estimate. Hence, the proposed channel predictor shall not only predict the channel, but also identify and equalize such errors. Once the entire resource grid has been equalized, either with estimated or predicted channel coefficients, data symbols are demapped and the resulting bits decoded.

\section{Channel Prediction with Hybrid CNN-LSTM}\label{sec:3_CNN}
\begin{figure}[t]
    \centerline{\includegraphics[width=7.5cm]{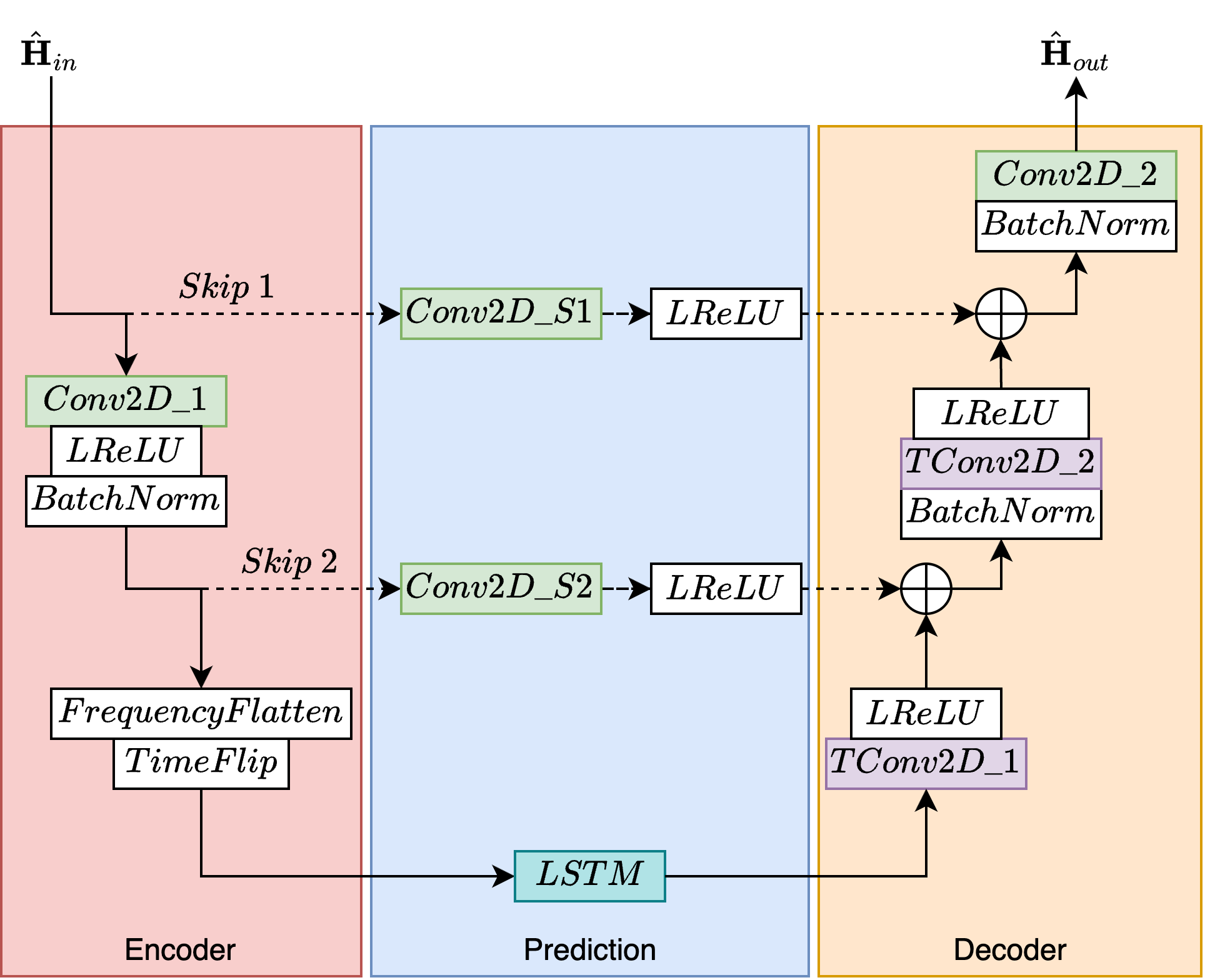}}
    \caption{Diagram of the proposed model.}
    \label{fig:CNNscheme}
\end{figure}
The proposed channel predictor's architecture, reported in Figure \ref{fig:CNNscheme}, comprises three sections. Its structure is partly inspired by the LinkNet architecture, an encoder-decoder NN with skip connections for semantic segmentation \cite{bib:linknet}, introducing temporal features extraction capabilities on top of it. First, complex features are extracted from the input channel matrix, maintaining the temporal duration of the grid and compressing its frequency span. This is achieved using a 2D convolutional (Conv2D) layer, which filters the input values using a set of pre-trained kernels; the specific parameters of this and the following layers are reported in Table \ref{tab:cnn}. Using the custom \textit{FrequencyFlatten} layer, the tensor is then flattened on the frequency axis, such that the remaining frequencies are regarded as additional channels. The \textit{TimeFlip} layer mirrors the resulting tensor on the temporal axis, ensuring that the first prediction depends on the most recent sample and not the most outdated one, resulting in improved prediction accuracy. In the second section, the obtained one-dimensional tensor is fed to an LSTM layer, exploiting its recurrent nature to extract temporal features from the compressed tensor to generate a time series of equal length. It must be noted that, at the same time, Conv2D layers on skip connections extract local temporal patterns from the input channel matrix and a less compressed representation, enhancing the features extracted by the LSTM layer. In the third and final section, the predicted CFR matrix is reconstructed by expanding the singleton frequency dimension to $N_{SC}$ subcarriers, adding at each stage the features extracted with skip connections. The frequency-specific features are generated using 2D transposed convolutional (TConv2D) layers, which apply the same kernels to the channels of each element of the input tensor. Finally, to smooth out the typical checkerboard pattern that characterizes TConv2D-based NNs, a Conv2D layer with linear activation function is employed as the last layer of the NN. Table \ref{tab:cnn} reports the parameters chosen for each trainable layer. We note that the padding (/cropping) vector for Conv2D (/TConv2D) layers reports the padding (/cropping) applied to the top, bottom, left, and right sides of the input tensor, in this order; "Same" implies that the layer employs enough padding (/cropping) to preserve the tensor's shape.
\setlength{\tabcolsep}{5pt}
\renewcommand{\arraystretch}{1.1}
\begin{table}[t]
    \centering
    \caption{Layers parameters}
    \label{tab:cnn}
    \begin{tabular}{|c|c|c|c|}
        \hline
        \textbf{Layer name} & \textbf{Filters/Units} & \textbf{Stride} & \textbf{Padding/Cropping}\\
        \hline
        Conv2D\_1  & 8 $(6\times3)$  & $[12, 1]$  & $[0, 0, 1, 1]$  \\
        \hline
        Conv2D\_S1 & 2 $(1\times3)$  & $[1, 1]$  & $Same$          \\
        \hline
        Conv2D\_S2 & 8 $(1\times3)$  & $[1, 1]$  & $Same$          \\
        \hline
        LSTM      & 16              & N/A       & N/A             \\
        \hline
        TConv2D\_1 & 8 $(4\times3)$  & $[1, 1]$  & $[0, 0, 1, 1]$  \\
        \hline
        TConv2D\_2 & 2 $(12\times3)$ & $[12, 1]$ & $[0, 0, 1, 1]$  \\
        \hline
        Conv2D\_2  & 2 $(3\times3)$  & $[1, 1]$  & $Same$          \\
        \hline
    \end{tabular}
\end{table}

\subsection{CNN-LSTM input}\label{ch:3.1_CNNin}
The estimated channel matrix $\hat{\mathbf{H}}_{est}$ is first normalized to unit average power; its real and imaginary components are then stacked together in an additional dimension, resulting in the real-valued input tensor $\hat{\mathbf{H}}_{in}$ of size $(N_{SC}, N_{sym}^{(slot)}, 2)$:
\begin{equation}
    \hat{\mathbf{H}}_{in} = \left[Re\left\{ \hat{\mathbf{H}}_{est}\right\}; Im\left\{ \hat{\mathbf{H}}_{est}\right\}\right].
\end{equation}
We note that, with this particular architecture, the polar representation of $\hat{\mathbf{H}}_{est}$ did not provide any loss improvement with respect to using the Cartesian representation during training.

\subsection{CNN-LSTM output}\label{ch:3.2_CNNout}
The CNN-LSTM model outputs a tensor with the same structure as the input tensor, \textit{i.e.}, $\hat{\mathbf{H}}_{out}$ is a tensor of size $(N_{SC}, N_{sym}^{(slot)}, 2)$, with the two channels containing the real and imaginary parts of the predicted CFR matrix.

\subsection{Dataset, training, and inference}\label{ch:3.3_DataTrainInfer}
A synthetic dataset of $N_{batch}$ estimated channel matrices and corresponding true following channel frequency responses is generated at each training epoch based on the described system model (Section \ref{ch:2_systemModel}), such that new data can be used to train the CNN-LSTM until the loss function convergence is reached. The chosen loss function to be optimized is the mean squared error, which coincides with the NMSE as a result of power normalization on the channel matrices. The training process employs a series of popular techniques, namely 1) L2 regularization, 2) early stopping, 3) learning rate warm-up, and 4) learning rate cosine annealing. After training, the model is employed for inference as is; online learning frameworks require an in-depth dedicated analysis, and are thus out of the scope of this work.
\setlength{\tabcolsep}{5pt}
\renewcommand{\arraystretch}{1.1}
\begin{table}[t]
    \centering
    \caption{Simulation parameters}
    \label{tab:parameters}
    \begin{tabular}{|c|c|}
        \hline
        \textbf{Parameter} & \textbf{Value} \\
        \hline
        Satellite altitude & 600 km \\
        \hline
        Training $E_b/N_0$ range & $E_b/N_0 = [0, 1, ..., 10]$ [dB] \\
        \hline
        Maximum Monte Carlo Iterations & $N_{MC} = 10^5$ \\
        \hline
        Data modulation order & $M=\{4, 16, 64\}$ \\
        \hline
        Pilot modulation order & $M_\pi = 4$ \\
        \hline
        Code rate & $R_c = 3/4$ \\
        \hline
        Channel models & \{NTN-TDL-A, NTN-TDL-C\} \cite{bib:tr38.811} \\
        \hline
        UE speed & $v_{UE}=\{5, 30, 50\}$ km/h \\
        \hline
        Carrier frequency & $f_c = 2$ GHz\\
        \hline
        Delay spread & $30$ ns\\
        \hline
        Number of subcarriers & $N_{SC} = 48$\\
        \hline
        Number of OFDM symbols & $N_{sym} = 28$\\
        \hline
        5G numerology & $\mu = 0$\\
        \hline
        Pilot indices per slot & $\mathcal{I}_\pi^{(slot)} = \left[3, 12\right]$\\
        \hline
        Batch size & $N_B = 1024$\\
        \hline
        Maximum learning rate & $0.03$\\
        \hline
        Minimum learning rate & $0.001$\\
        \hline
        Learning rate warm-up duration & $40$ epochs\\
        \hline
        Cosine annealing period & $100$ epochs\\
        \hline
        Early stopping patience & $3$ cosine annealing cycles\\
        \hline
        L2 regularization factor & $10^{-6}$\\
        \hline
    \end{tabular}
\end{table}

\section{Results}\label{ch:4_results}
We assess the performance of the proposed predictor in an end-to-end simulation in the MATLAB computing environment, where a new resource grid is generated at each Monte Carlo iteration. The simulation parameters are reported in Table \ref{tab:parameters}; we note that, except where stated otherwise, the predictor employed in such tests has been trained on 16-QAM data symbols under NTN-TDL-C channel model \cite{bib:tr38.811}, with a UE speed of 5 km/h. We set $\sigma_D = \frac{0.1\cdot 10^{-6}\cdot f_c}{3}$, such that the residual frequency synchronization error magnitude does not exceed the 5G requirement of 0.1 parts per million of the carrier frequency 99.7\% of the times \cite{bib:sync} ($3\sigma$ rule). The performance of the proposed algorithm are compared to those of a channel-estimation-based system with pilot-full slots only.

\subsection{Bit Error Rate}\label{ch:4.1_BER}
\begin{figure}[t]
    \centerline{\includegraphics[width=7cm]{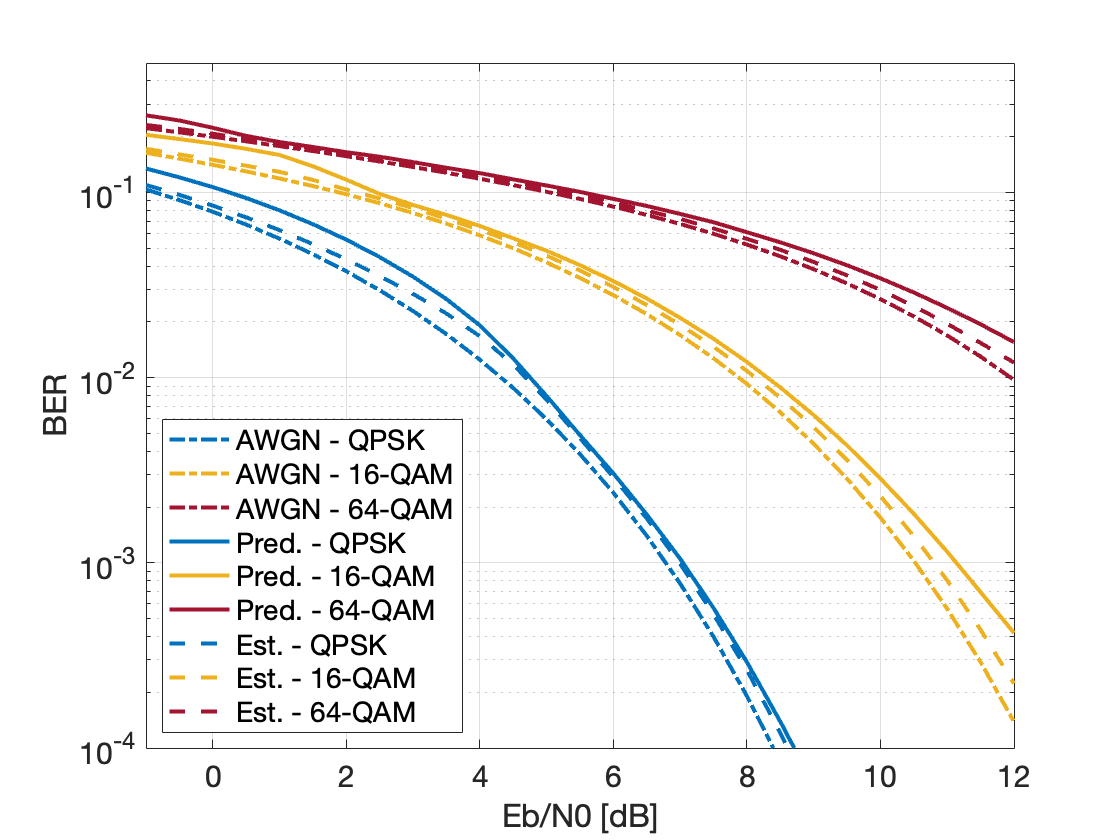}}
    \caption{BER as a function of $E_b/N_0$ (NTN-TDL-C).}
    \label{fig:ber}
\end{figure}
We first report the uncoded BER for different QAM modulation orders as a function of the $E_b/N_0$ (Figure \ref{fig:ber}). For each modulation order, three regions can be identified in the plot. At low $E_b/N_0$, the channel prediction performances are hindered by the noisy input channel estimates; \textit{e.g.}, with 16-QAM modulation the performance gap is more than 0.5 dB of $E_b/N_0$ until the 10\% BER threshold is reached. As the average energy per bit increases, the prediction accuracy becomes comparable with that of channel estimation, leading to a second region with overlapping BER curves. Finally, at high $E_b/N_0$ channel prediction is limited by the algorithm's accuracy, which depends on the specific architecture and training procedure: this typically results in a BER error floor, which here lies outside of the maximum assessed $E_b/N_0$ value, proving the effectiveness of the proposed model. Indeed, the channel prediction BER for quadrature phase shift keying (QPSK)-modulated data symbols reaches $10^{-4}$ at 8.5 dB of $E_b/N_0$, approximately the same as the channel estimation BER: this is due to the fact that only phase inaccuracies affect the QPSK demapper; on the opposite, the smaller inter-symbol distance in the 64-QAM constellation results in a wider the gap between channel prediction and estimation at high $E_b/N_0$, with the prediction achieving a 1.5\% BER at 12 dB (a 0.5 dB loss with respect to channel estimation). 

\subsection{Throughput}\label{ch:4.2_throughput}
\begin{figure}[t]
    \centerline{\includegraphics[width=7cm]{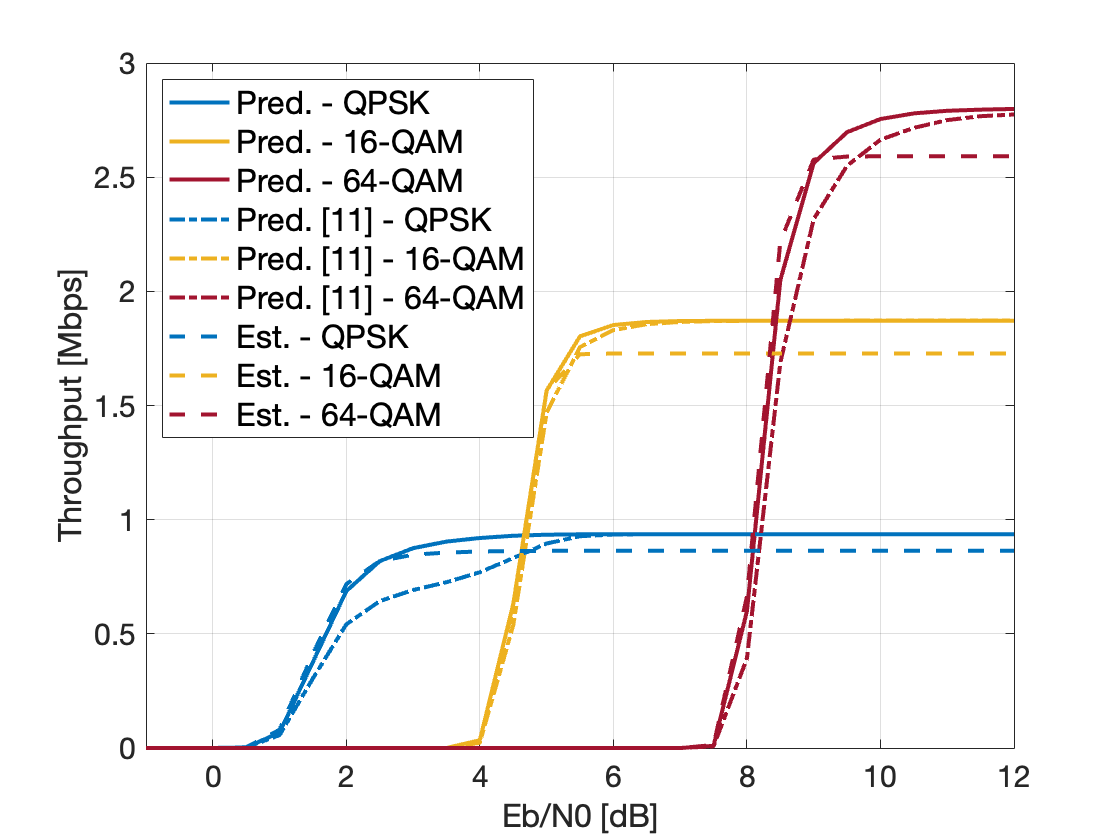}}
    \caption{Throughput as a function of $E_b/N_0$ and QAM modulation order (NTN-TDL-C, $v_{UE}=5km/h$).}
    \label{fig:tp}
\end{figure}

We showed that the proposed algorithm introduces only a limited amount of bit errors in the demodulated bit sequences; however, such errors may in turn hinder the BLER and, thus, the TP gains that can be achieved with the removal of pilot symbols. Starting from the BLER, assessed as the ratio between the number of received block errors over the total number of transmitted blocks, we evaluate the effective TP in the estimation-based and prediction-based systems as:
\begin{gather}
    TP_{e} = \frac{m \cdot R_c \cdot N_{SC} \cdot (N_{sym}^{(slot)} - \left| \mathcal{I}_\pi^{(slot)}\right|)}{T_{slot}}\cdot(1-BLER_{e}), \\
    TP_{p} = \frac{TP_e + \frac{m \cdot R_c \cdot N_{SC} \cdot N_{sym}^{(slot)}}{T_{slot}}(1-BLER_{p})}{2},
\end{gather}
respectively. $T_{slot}=1ms$ represents the duration of an OFDM slot, while $m=log_2M$ is then number of bits per data symbol. We report in Figure \ref{fig:tp} the TP achieved using various data modulation orders in the NTN-TDL-C channel as a function of the $E_b/N_0$; we also include the results obtained in \cite{bib:icmlcn} as a benchmark.
Regardless of the data modulation order, at low $E_b/N_0$ the proposed channel predictor provides TPs on par with those offered by channel estimation: indeed, the two curves approximately overlap on their rising sections until the peak throughput for channel estimation is reached. The TP gains achieved through prediction appear after an $E_b/N_0$ of 2.5 dB, 5 dB, and 9 dB, with QPSK, 16-QAM, and 64-QAM, respectively. We note that the proposed architecture achieves substantial gains with respect to the Conv2D/TConv2D-based model presented in \cite{bib:icmlcn}: the introduction of a LSTM layer dedicated to prediction greatly improves the low $E_b/N_0$ performance, with the QPSK TP exceeding the channel estimation performance with a 2.5 dB gain with respect to the benchmark model. More in general, for all of the considered data modulation orders, the proposed model requires a lower $E_b/N_0$ to exceed the channel estimation performance and reach the peak TP (0.94 Mbps for QPSK, 1.87 Mbps for 16-QAM, and 2.81 Mbps for 64-QAM). However, such findings may not translate well to different channel conditions, as the predictor has been trained on a dataset with a fixed Doppler spread and NTN-TDL-C channel model. For this reason, we further test the predictor to analyze the impact of dataset mismatch.

\subsection{Throughput with Doppler spread mismatch}\label{ch:4.3_throughput_speed}
\begin{figure}[t]
    \centerline{\includegraphics[width=7cm]{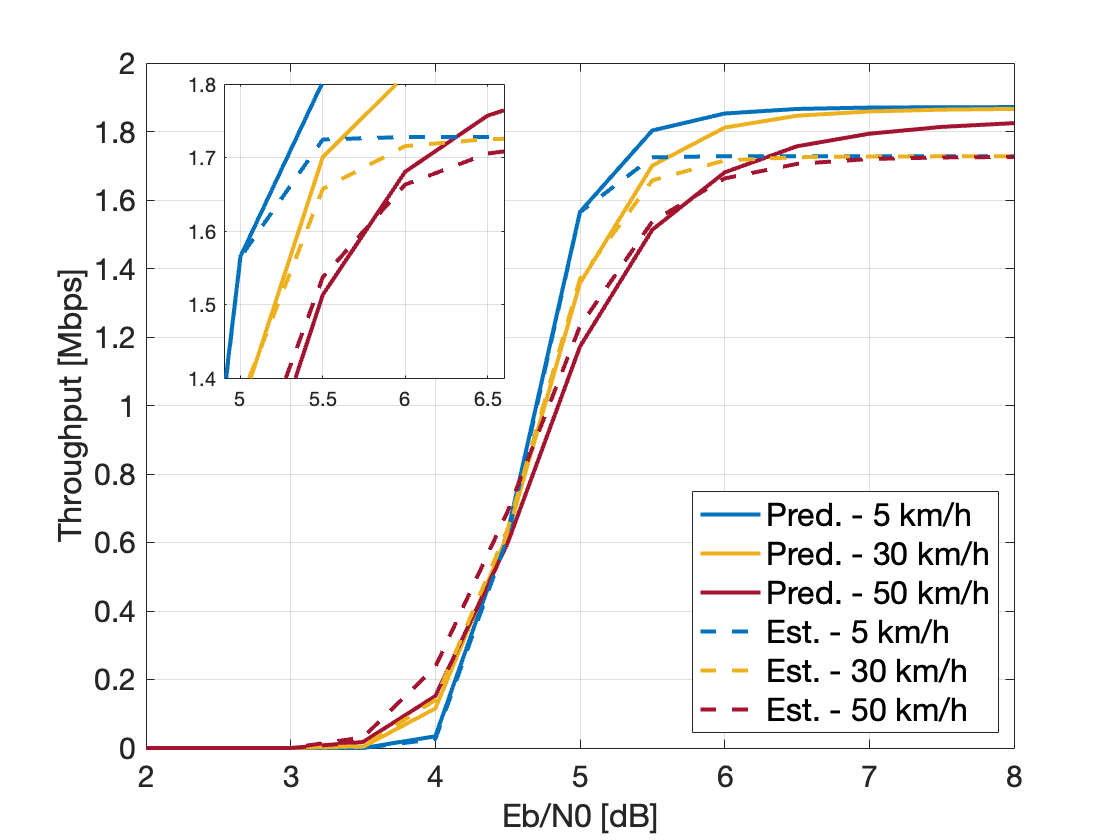}}
    \caption{Throughput as a function of $E_b/N_0$ and UE speed (NTN-TDL-C, $M=16$).}
    \label{fig:tp_speed}
\end{figure}

As previously mentioned, it is imperative to ensure that the model's capabilities generalize well to the velocities and, thus, Doppler spreads that can be expected in real world scenarios. Clearly, with higher UE speeds more dynamic channel temporal patterns can be expected, making accurate predictions more difficult to achieve. Figure \ref{fig:tp_speed} shows that at 30 km/h, corresponding to, \textit{e.g.}, electric scooters, the prediction TP overcomes estimation at 5 dB of $E_b/N_0$, while the peak TP of 1.87 Mbps is reached at 8 dB $E_b/N_0$. Even at 50 km/h, often the speed limit in urban areas, channel prediction provides benefits to the TP starting from an $E_b/N_0$ of 5.75 dB; however, at such a high speed minor performance degradation is experienced at lower $E_b/N_0$, \textit{e.g.}, the predictor achieves 1.17 Mbps of TP at 5 dB against 1.23 Mbps obtained with estimation. Nonetheless, it is clear that the proposed model did not overfit on the training Doppler spread. However, we note that a more complex model may be required to achieve accurate predictions with higher UE speeds.

\subsection{Throughput with channel model mismatch}\label{ch:4.4_throughput_channel}
\begin{figure}[t]
    \centerline{\includegraphics[width=7cm]{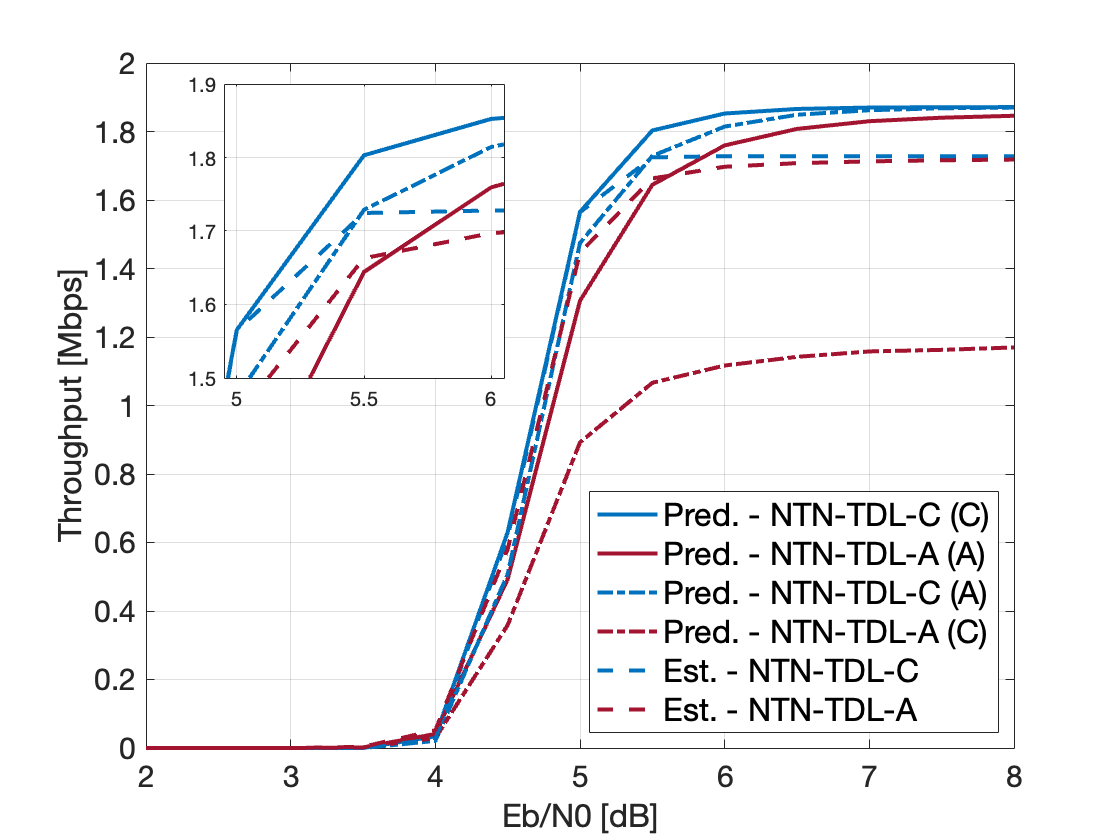}}
    \caption{Throughput as a function of $E_b/N_0$ for different channel models (labeled as "test (training)") ($M=16$, $v_{UE}=5$ km/h).}
    \label{fig:tp_channels}
\end{figure}

We further assess the performance of the proposed predictor on different channel models. In this analysis, we train a second neural network with the same architecture on the NTN-TDL-A channel model. Figure \ref{fig:tp_channels} reports the TP results, where each predictor is tested on both the NTN-TDL-C and -A channel models; the legend reports the test channel model, followed by the training channel model in round brackets. Focusing on the tests on the NTN-TDL-A channel model (red curves), the matching predictor (continuous line) provides TP improvements with respect to channel estimation starting from approximately 5.5 dB of $E_b/N_0$, with 99\% of the peak TP being reached at 8dB; on the opposite, the mismatched predictor (dash-dotted line) cannot provide satisfying performance, reaching only 1.2 Mbps at the same $E_b/N_0$ level. Moving to the NTN-TDL-C tests (blue curves), it is clear that the same does not apply to the dual case. Indeed, the model trained on NTN-TDL-A (dash-dotted curve) provides excellent performance when tested on the NTN-TDL-C channel model, overcoming the channel estimation performance with just a 0.5 dB loss with respect to the matched predictor. This observation suggests that the CNN-LSTM learns to rely on the prediction of the most important, \textit{i.e.}, the most powerful, taps in the considered channel model. When the training dataset includes a LoS component, the predictor relies on such component for the prediction; thus, when trained on the non-LoS (NLoS) channel model, the predictor is able to generalize fairly well to different channel conditions and does not necessitate online training techniques to overcome the channel estimation performance.

\section{Computational complexity}\label{ch:5_complexity}
We conclude the analysis of the proposed predictor with a discussion on its computational complexity. We here focus on the number of multiplications carried out in the layers of the proposed predictor as an upper bound to multiply and accumulate (MAC) units. Assuming that the $i$-th Conv2D or TConv2D layer has $C^{(i)}$ kernels of shape $(W_f^{(i)}, W_t^{(i)})$ and that its input tensor has shape $(L_f^{(i)}, L_t^{(i)}, N^{(i)})$, the corresponding MACs can be upper bounded by:
\begin{gather}
    Mul_{i}^{(Conv2D)} = L_f^{(i+1)}L_t^{(i+1)}N^{(i)}W_f^{(i)}W_t^{(i)}C^{(i)}, \\
    Mul_{i}^{(TConv2D)} = L_f^{(i)}L_t^{(i)}N^{(i)}W_f^{(i)}W_t^{(i)}C^{(i)}.
\end{gather}
On the other hand, an LSTM layer with $U^{(i)}$ hidden units and as many cells as the input/output sequences length $L_t^{(i)}$ performs the following amount of multiplications \cite{bib:lstmComplexity}:
\begin{equation}
    Mul_{i}^{(LSTM)} = L_t^{(i)}U^{(i)}(4N^{(i)}+4U^{(i)}+3).
\end{equation}
Considering Table \ref{tab:cnn}, the number of MACs required for inference with the proposed model is approximately 157k (neglecting the contributions from BN and LReLU), while the number of trainable parameters is 5.8k. In comparison, the benchmark presented in \cite{bib:icmlcn} has similar complexity (160k MACs, 5.5k trainable parameters), but provides far inferior performance. Low power hardware accelerators, which may be suitable for implementation on board of LEO satellites, have been proven able to run CNNs at 100+ inferences per second, \textit{e.g.}, \cite{bib:hardware} reports that a 1.1M MACs CNN with 13 sequential Conv2D layers and a fully-connected layer was run on a nano unmanned aerial vehicle at 139 inferences/s (an inference latency of 7 ms) with 100 mW of power. While our algorithm requires far lower MACs, it includes a larger number of sequential operations in the LSTM layer. Thus, despite its low complexity, the real-time feasibility of the CNN-LSTM requires a more in-depth analysis, possibly including an in-hardware implementation, which is out of the scope of this work. Nonetheless, scaling down the inference latency of \cite{bib:hardware} based on the MACs of their and our model results in $\sim$1 ms/inference, which coincides with the OFDM slot duration: a promising result in the direction of real-time prediction.

\section{Conclusions}
This paper presented a CNN-LSTM-based channel predictor for uplink communications in NTNs. The DL model employs Conv2D and TConv2D layers to, respectively, compress and expand in the frequency axis the CFR matrix estimated over an OFDM slot, while the temporal prediction is carried out on the compressed matrix by an LSTM layer. We show that the proposed predictor allows pilots to be removed every other OFDM slot, resulting in an increase in the peak uplink throughput with virtually no performance degradation at any $E_b/N_0$ level with various data modulation order. The CNN-LSTM is resilient to mismatches between training and test UE speeds as long as they are under 50 km/h. We further showed that, when trained on the NLoS NTN-TDL-A channel model, the predictor was able to provide good throughput improvements on both the NTN-TDL-A and the NTN-TDL-C model, losing only 0.5 dB of $E_b/N_0$ on the latter with respect to matched training and test channel models. The proposed CNN-LSTM is lightweight, requiring only 157k MACs and having just 5.8k trainable parameters; however, due to the large number of sequential operations, its real-time feasibility must be assessed in a dedicated analysis, possibly including an in-hardware implementation. Further studies should also investigate strategies to extend the predictor to different frequency spans, \textit{e.g.}, with separate predictions on a single physical resource block of 12 subcarriers, and the extension to multiple input and output OFDM slot to obtain more accurate predictions and/or further pilot reductions.

\section{Acknowledgments}\label{Acknowledgment}
This work was partially supported by the European Union under the Italian National Recovery and Resilience Plan (NRRP) of NextGenerationEU, partnership of "Telecommunications of the Future" (PE00000001 - program "RESTART"), and by the 6G-NTN project, which received funding from the Smart Networks and Services Joint Undertaking (SNS JU) under the European Union’s Horizon Europe research and innovation programme under Grant Agreement No 101096479. The views expressed are those of the authors and do not necessarily represent the project. The Commission is not liable for any use that may be made of any of the information contained therein.

\end{document}